# Polarization control and sensing with two-dimensional coupled photonic crystal microcavity arrays

Hatice Altug[*] and Jelena Vučković[†]

*Edward L. Ginzton Laboratory, Stanford University, Stanford, CA 94305-4088*

## Abstract

We have experimentally studied polarization properties of the two-dimensional coupled photonic crystal microcavity arrays, and observed a strong polarization dependence of the transmission and reflection of light from the structures - the effects that can be employed in building miniaturized polarizing optical components. Moreover, by combining these properties with a strong sensitivity of the coupled bands on the surrounding refractive index, we have demonstrated a detection of small refractive index changes in the environment, which is useful for construction of bio-chemical sensors.

[*] Also at the Department of Applied Physics, Stanford University, Stanford, CA 94305, altug@stanford.edu
[†] Also at the department of Electrical Engineering, Stanford University, Stanford, CA 94305, jela@stanford.edu, http://www.stanford.edu/group/nqp

We have recently proposed [1] and experimentally demonstrated [2] two-dimensional coupled photonic crystal resonator arrays (2D CPCRAs), exhibiting a small group velocity over all wavevectors and in all crystal directions. These structures are interesting for construction of low-threshold devices such as photonic crystal (PhC) lasers with increased output powers and various nonlinear optical components. In our prior experimental work, we measured the band diagram of such a structure by testing transmission through it at various incidence angles, thereby controlling the in-plane wavevector (k-vector); we demonstrated a small group velocity (below 0.008c at the Γ point) for a broad range of k-vectors [2]. In this article, we demonstrate a strong sensitivity of the light transmitted and reflected from CPCRAs on the input polarization and on the surrounding refractive index, which can be employed in building miniaturized polarizing optical components or bio-chemical sensors.

When PhC microcavities are tiled in two dimensions, thereby forming a 2D CPCRA (as shown in Fig. 1b), defect modes of individual cavities form coupled bands located inside the photonic bandgap of the surrounding photonic crystal. In particular, coupled arrays in a square PhC lattice exhibit three coupled bands: monopole, dipole, and quadrupole [1-2]. Here we focus on the coupled dipole band, which is (in a structure with a fourfold rotational symmetry) doubly degenerate at the Γ point, but splits into two sub-bands (x- and y-dipoles) in the ΓX direction [1]. The electromagnetic fields of the two dipole modes are related by 90° rotation and field components of the x-dipole are shown in Fig. 2a; the dominant electric field components of the x- and y-dipole modes are $E_x$ and $E_y$, respectively, and one can employ the input field polarization to preferentially excite one or the other mode. Designed CPCRAs were fabricated in silicon on insulator (SOI) by employing the procedure described in Reference 2, and SEM pictures are shown in Fig. 1b. The size of the array is 100μm by 100μm, and the PhC parameters are the periodicity $a$=488 nm, the hole radius $r$ = 190nm, and the slab thickness $d$=275 nm. The coupled cavities have two PhC layers between them in all directions and the unit cell size is $3a \times 3a$, as shown in the inset of Fig. 3. For this set of parameters, the Finite-Difference Time-Domain (FDTD) method predicts that the dipole mode is in the scanning range of our tunable laser (1460nm-1580nm). Our experimental setup is shown in Fig. 1a. The structure is excited at the vertical incidence (in the z-direction, i.e., at the Γ point of the band diagram) by a tunable laser whose beam is linearly polarized in the x-direction and is slightly focused on the sample by a very low numerical aperture (NA) objective lens. A total transmitted signal through the structure, as well as reflected signals of the same (x) or opposite (y) polarization are measured (the reflected signals are separated by employing a polarizing beam splitter in front of the top detector). The rotation of the structure around the z-axis is used to test the polarization dependence of the reflection and transmission. After the structure is rotated by an angle $\phi$ around the z-axis, the excitation beam (x-axis) is thus polarized in the direction $\phi$ of the structure, and the reflected signal of opposite polarization (y-axis) then corresponds to a linearly polarized signal in the direction $\phi+90^o$ of the structure. The excitation of the high symmetry directions $\Gamma X$ and $\Gamma M$ hence corresponds to $\phi$ equal to $0^o$ and $45^o$, respectively; the x-dipole is expected to be primarily excited for $\phi=0^o$, and the y-dipole for $\phi=90^o$.

Fig. 2b shows the transmission through the structure at vertical incidence ($\Gamma$ point), and for the input polarization in the $\phi=0^o$ and $\phi=90^o$ directions. The dips in the red and blue curves correspond to the x- and y-dipole modes with wavelengths 1564 nm and 1555 nm, respectively; this implies that the degeneracy of the dipole band in our structure is lifted at the $\Gamma$ point, as a result of the structural asymmetry under $90^o$ rotation (holes are slightly elliptical, i.e., their radius in the x-direction is roughly 5% larger than in the y-direction) [2]. The reduction of rotational symmetry induces strong polarization effects and can be employed to construct polarization sensitive optics. As can be seen in Fig. 2b, for an input beam at the wavelength of the x-dipole mode (1564 nm) vertically incident to the structure, the y-component of the beam will be transmitted, while the x-component will be strongly reflected. Similarly, if the input beam has the wavelength of the y-dipole mode (1555 nm), then the previously described transmitted and reflected components will be interchanged. This implies that such structures can be used as miniaturized polarizing mirrors. It is clear that PhC structures with a 90-degree rotational symmetry (i.e. with circular holes) have degenerate x- and y- dipole modes; hence, they are polarization insensitive and cannot be used as polarizing optics. Recently, another group has theoretically analyzed a polarizing mirror effect in an asymmetric square PhC lattice [3]. The advantage of asymmetric CPCRAs for this application is that, because of the employed flat bands, they behave as polarizing mirrors even for the input beam that is not vertically incident. This is also useful for making polarizing beamsplitters. Moreover, the described effect in our structures can also be employed to control the output polarization of a CPCRA laser that we proposed recently [1,2].

An experimental test of whether a device is indeed a polarizing mirror can be performed by measuring reflection from it for varying linear polarizations of the incident beam. This is similar to the test of a linear polarizer: for an input beam polarized in the direction θ relative to the transmission axis of the polarizer, the transmitted intensity should be proportional to $\cos^2(\theta)$, which is a well known Malus's Law [4]. Since we measure the reflected signal at opposite polarization (φ+90°) relative to the vertically incident excitation (polarization angle φ), it can be shown (by a simple geometrical argument) that the reflected intensity at opposite polarization should be proportional to $\sin^2(2\phi\pi/180)$. The reflected signal is minimum for an incident beam polarized along ΓX direction (φ=0°, 90°), and maximum for the incident polarization along ΓM direction (φ=45°), as experimentally observed (Figs. 2c and 3). The physical explanation for the observed effects is as follows: for the input polarization angle φ=0°, 90°, one of the two dipole modes is preferentially excited, and it is expected from Fig. 2b that the reflection is maximized and polarized same as the excited mode, i.e., primarily in the direction φ (the reflected signal at the opposite polarization φ+90° should thus be minimized). On the other hand, for the input polarization φ=45°, both x- and y-dipole modes can be excited; since both of them have non-zero field components in the direction φ=135°, the reflected signal at the opposite polarization is strong. As seen in Fig. 2c, the x-dipole peak is more pronounced for φ=45°, as a result of its higher Q factor in this particular structure, which is clear from the narrower width of transmission in Fig. 2b. 3D FDTD simulation of the studied structure has also confirmed this difference in the Q-factors of the dipole modes. It should also be noted in Figs. 2c and 3 that the measured reflected signal with opposite polarization for φ=0°, 90° is very weak, but is not zero. This weak signal results from the fact that the x- and y-dipole modes are not purely linearly polarized (see Fig. 2a). Therefore, an input beam polarized in the x- (y-) direction can also weakly excite the y- (x-) dipole mode at the resonance wavelength of the mode. When this happens, the reflection is polarized as the excited mode, which is in this case primarily opposite to the polarization of the excitation; this process is referred to as polarizing mixing. For example, a small peak at the wavelength of the y-dipole can be observed in the reflected signal with opposite polarization ($\phi=90$) for input polarization $\phi=0^o$ (Fig. 2c). Similar polarization control has been recently observed in elliptic metal nanohole arrays, but with broader peaks, resulting from losses in metals and low Q-factors of surface plasmon modes [5]; CPCRAs minimize non-radiative absorption losses and can achieve narrow bandwidth polarization effects for wide range of incidence angles (resulting from the flat bands).

A strong reflected signal with opposite polarization for $\phi = 45^o$ is a result of interaction of light with CPCRA and carries information mostly originating from the structure. In fact, we performed the same experiment on a SOI slab without PhC, and measured the signal of opposite polarization which was 50 times weaker than the one shown in Fig. 2c for $\phi = 45^o$, while none of the described resonances were observed. A very small scattering of the input light into opposite polarization on SOI slab could happen as a result of an irregularity at the structure surface, but this is a much weaker signal than the one from CPCRA. In CPCRA, the position of the resonance in the reflected signal strongly depends on the surrounding refractive index, and it can be employed in high Signal-to-Noise Ratio (SNR) bio-chemical sensing. In order to demonstrate the detection of the refractive index change, we use another CPCRA with a=467nm, r=195nm, and d=255nm, in which the dipole modes are shifted to shorter wavelengths, so that they remain in the wavelength range of our tunable laser even after the structure is immersed in an environment with increased refractive index. The y- and x-dipole resonances are 1525nm and 1536nm, respectively, and the y-dipole mode has a higher Q in this case. The reason for this is the difference is in the way asymmetry is introduced in this structure; as opposed to the first tested sample, the holes in this structure are not elliptical, but the periodicity is different in the x and y directions. The periodicity in the x-direction is closer to the designed, so the y-dipole mode has a better Q. This can also be confirmed from the widths of the transmission spectra (similar to Fig. 2b) and from the 3D FDTD simulations. Sensing of refractive index changes is performed by dropping a small amount of isopropanol (IPA, refractive index n=1.377) and methanol (n=1.328) separately on the same structure. The reflected signal of opposite polarization for the structure covered with IPA is shown in Fig 4c: the spectrum is clearly shifted to longer wavelengths, as expected. The reflection spectrum from the same structure for methanol is very similar to IPA, again with two peaks, but at shorter wavelengths relative to IPA, as expected. Before using methanol, the structure was cleaned and the reference signal repeated. We have estimated the wavelength shift theoretically by the FDTD method for the structure without any asymmetry. Since the size of the droplet (diameter ~5mm) is large compared to the structure, we have assumed in the simulation that the liquid completely surrounds CPCRA and penetrates holes. Fig. 4a shows the resonance wavelength shift as a function of the surrounding refractive index, and a good agreement between theory and experiment can be observed (the two experimental lines correspond to the x- and y-dipoles). By immersing the structure into IPA instead of methanol, the refractive index changes by $\Delta n = 0.049$, and the dipole band position shifts by $\Delta\lambda = 7$nm. 3D FDTD simulation predicts 0.521nm shift in wavelength for an index change of $\Delta n = 0.002$, a sensitivity similar to that observed recently in a single PhC cavity sensor in the passive configuration [6]. The CPCRA approach simplifies the positioning of the analyzed material, since the cavities are distributed over a larger area, and it enables a high sensitivity at different incidence angles due to flat-coupled bands (alignment tolerance). The sensitivity can be improved by changing the type of cavities inside CPCRAs, or by embedding an active quantum well layer inside the structure and using it above lasing threshold, where the linewidth narrows [7].

In conclusion, we have experimentally analyzed polarization properties of 2D CPCRAs, and observed a strong dependence in the transmitted and reflected signal on the input polarization. We have confirmed that the structures can be used as miniaturized polarizing optics: mirrors and beamsplitters. Finally, we have combined the detection of the reflected signal from the structure with a strong sensitivity of CPCRAs on the surrounding refractive index to demonstrate detection of refractive index changes of 0.05. Our theoretical analysis shows that this can be improved by at least an order of magnitude, which is interesting for construction of bio-chemical sensors.

**Acknowledgment:** This work has been supported by the Marco Interconnect Focus Center, and in part by the MURI Center for Photonic Quantum Information Systems (ARO/ARDA program DAAD19-03-1-0199), the Charles Lee Powell Foundation Faculty Award and the Terman Award.

# Figure Captions:

**Figure 1.** (a) Experimental setup used in testing of the polarization properties CPCRA and in sensing. The explanation of symbols is as follows: $\lambda/2$ - half-wave plate, BS – non-polarizing beam splitter, PBS - polarizing beam splitter, OL - objective lens, IR-cam - infrared camera and D – detector. Rotation of the structure around the z- axis is controlled by a rotation stage. (b) SEM pictures of the fabricated CPCRA with $A=3a$.

**Figure 2. (a)** Electromagnetic field pattern of the coupled x-dipole band at the $\Gamma$ point and at the center of the PhC slab (Field components of the y-dipole are obtained by first interchanging $E_x$ and $E_y$, and then by rotating all three patterns by 90º) **(b)** Transmission spectra of the CPCRA for the coupled x- (red, $\phi=0º$) and y- (blue, $\phi=90º$) dipole modes at the $\Gamma$ point. **(c)** Reflection spectra of the CPCRA at the opposite polarization relative to the excitation, for $\phi=0º$, 45º and 90º respectively.

**Figure 3.** The power of the reflected signal with opposite polarization relative to the excitation as a function of the input polarization angle $\phi$ (circles). The solid line is a sine square function fit to the experimental data. The inset shows a unit cell of CPCRA with $A=3a$ and high symmetry directions.

**Figure 4 (a)** The resonance wavelength as a function of the surrounding refractive index. The three data points correspond to the structure in air (n=1), methanol (n=1.328) and IPA (n=1.377). Triangles and stars correspond to the experimental wavelength shifts of the y-dipole and x-dipoles respectively, and circles show the wavelength shift simulated by FDTD for the same structure without any asymmetry. Lines are fitted to experimental data points. **(b)** The reflection spectrum with opposite polarization with respect to the excitation for the structure in air. Arrows indicate the positions of the x- and y- dipole bands **(c)** The same spectrum when the structure is immersed in IPA.

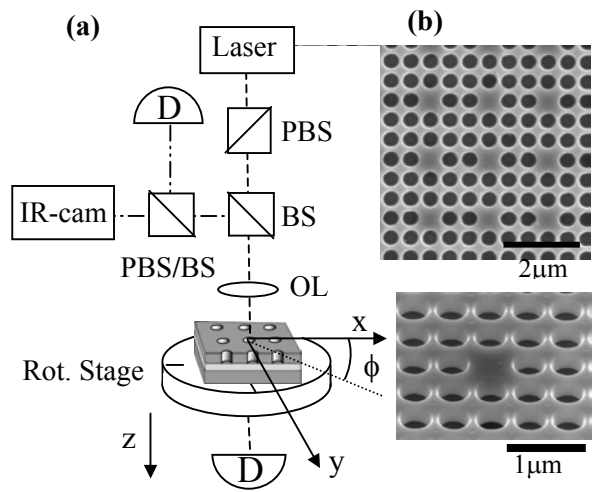

Figure 1

Authors: H. Altug, J. Vuckovic

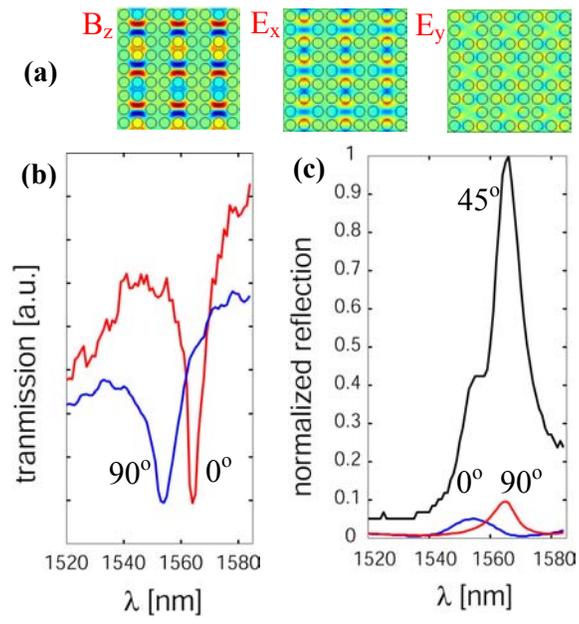

Figure 2

Authors: H. Altug, J. Vuckovic

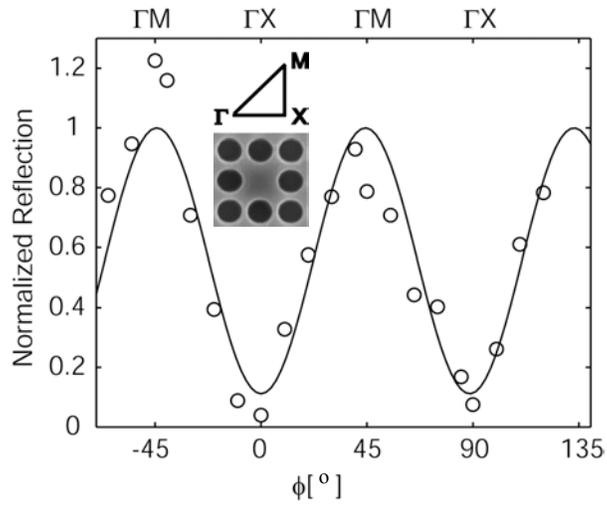

Figure 3

Authors: H. Altug, J. Vuckovic

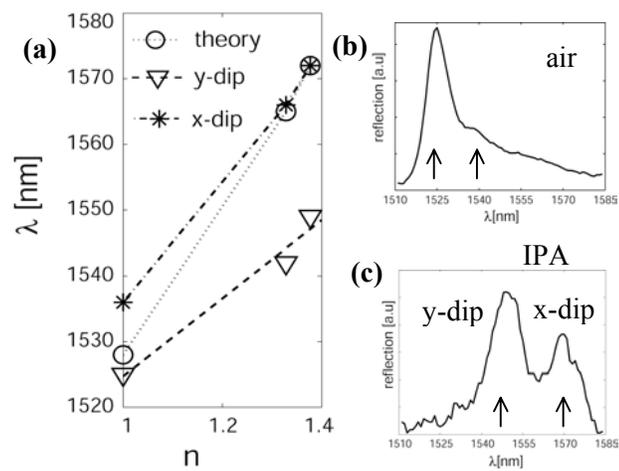

Figure 4

Authors: H. Altug, J. Vuckovic